\documentclass[preprint2]{aastex} 

\usepackage{amsmath}
\usepackage{color}

\newcommand{\half}{\ensuremath{{\textstyle \!\frac{1}{2}}}}

\shorttitle{First stereoscopic coronal loop reconstructions}
\shortauthors{Feng et al.}

\begin{document}

\title{First stereoscopic coronal loop reconstructions from
            STEREO/SECCHI images}

\author{L. {Feng}\altaffilmark{1,2},
        B. {Inhester}\altaffilmark{1},
        S. {Solanki}\altaffilmark{1},
        T. {Wiegelmann}\altaffilmark{1},
        B. {Podlipnik}\altaffilmark{1},
        R.A. {Howard}\altaffilmark{3}, and
        J.-P. {Wuelser}\altaffilmark{4}}
\date{DOI: 10.1086/525525 \\
Bibliographic Code: 2007ApJ...671L.205F}
\altaffiltext{1}{Max-Planck-Institut f\"ur Sonnensystemforschung,
       Max-Planck-Str.2,37191 Katlenburg-Lindau, Germany}
\altaffiltext{2}{Purple Mountain Observatory,
  Chinese Academy of Sciences, Nanjing, China}
\altaffiltext{3}{Naval Research Laboratory, Code 7660, 4555 Overlook Ave. SW,
  Washington D.C., USA, 20375}
\altaffiltext{4}{Solar and Astrophysics Lab., Lockheed Martin ATC,
  3251 Hanover St., Palo Alto, CA 94304, USA}

\begin{abstract}

  We present the first reconstruction of the three-dimensional shape
  of magnetic loops in an active region from two different vantage
  points based on simultaneously recorded images.
  The images were taken by the two EUVI telescopes of the SECCHI
  instrument onboard the recently launched STEREO spacecraft when the
  heliocentric separation of the two space probes was 12 degrees.
  We demostrate that these data allow to obtain a reliable
  three-dimensional reconstruction of sufficiently bright loops. The
  result is compared with field lines derived from a coronal magnetic
  field model extrapolated from a photospheric magnetogram
  recorded nearly simultaneously by SOHO/MDI.
  We attribute discrepancies between reconstructed loops
  and extrapolated field lines to the inadequacy of the linear force-free
  field model used for the extrapolation.

\end{abstract}


\section{Introduction}

With the launch of NASA's STEREO mission in October 2006, a new
dimension of solar coronal observations has been opened. For the first
time, objects above the solar surface can be perceived in three
dimensions by analysing the stereo image pairs observed with the
SECCHI instruments onboard the STEREO spacecraft and without making
a-priori assumptions about their shape.
The two STEREO spacecraft orbit the Sun at approximately 1 AU near the
ecliptic plane with a slowly increasing angle of about 45 degrees/year
between STEREO A and STEREO B. Each spacecraft is equipped with,
among other instruments, an EUV telescope (SECCHI/EUVI).
For the objectives of the mission and more details about the
EUVI telescopes see \citet{Wuelser:etal:2004} and \citet{Howard:etal:2007}.

The major building blocks of the solar corona are loops of magnetic
flux which are outlined by emissions at, e.g., EUV wavelengths. In
principle, the magnetic field in the lower corona can be derived from
surface magnetograms by way of extrapolations
\citep[e.g.][]{wiegelmann:2007}. However, missing boundary values and
measurement errors may introduce considerable uncertainties in the
extrapolation results so that there is an obvious need for an
alternative three-dimensional determination of the
coronal magnetic field geometry. Among other goals of the mission,
this requirement has been one of the drivers for STEREO.

Attempts for a three-dimensional reconstruction of the coronal
magnetic field from EUV observations have started long before STEREO
data was available and date back more than a decade
\citep{Berton:Sakurai:1985,Kouchmy:Molodensky:1992}.
Here, we for the first time use two simultaneously observed EUVI
images observed by the two STEREO probes and rigourously reconstruct
loop shapes without any further assumption about their temporal or
spatial behaviour from which earlier reconstructions employing consecutive images from
a single spacecraft suffered \citep{aschwanden:etal:2006}.
We compare the reconstruction results with field lines derived
from linear force-free magnetic field models with variable $\alpha$,
the ratio of field-aligned current density to field strength
\citep{seehafer:1978}.

\section{The data}

\begin{table}
\begin{tabular}{lrr}
STEREO probe                         &  A  & B \\ \hline
Helioc. dist. (AU)       &  1.068788 &  0.958071 \\
Sun's app. rad. (arcsec) &   897.866 &  1001.625 \\
Longitude (degrees) &  -4.277 & 7.524 \\
Latitude (degrees)  &  -0.293 & 0.095 \\ \hline
\end{tabular}
\caption{STEREO spacecraft coordinates at the time of the observations.
Spacecraft longitude and latitude are given in the Heliocentric Earth
Ecliptic (HEE) coordinate system.}
\label{tab:orbitpos}
\end{table}

For our reconstruction we used
EUV images at $\lambda$ = 17.1 nm taken by the almost identical
SECCHI/EUVI telescopes onboard of the two STEREO spacecraft at
2007-06-08 03:21 UT when the well isolated active region NOAA 0960 was
close to solar disk centre. The line $\lambda$ = 17.1 nm is
emitted by the Fe IX ion which in thermal equilibrium forms at about
1.1 million K.
At the time of these observations, the two STEREO spacecraft had a
heliocentric separation of 11.807 degrees. The precise spacecraft
positions at the time of the observation are listed in table
\ref{tab:orbitpos}

For a comparison of our reconstruction with magnetic field lines we
made use of a SOHO/MDI magnetogram \citep{scherrer:etal:1995} taken only
9 seconds prior to the EUVI images.
The active region is well isolated from neighbouring field sources so
that an extrapolation of the surface field is possible.
MDI, however,  provides only the line-of-sight field component, which
for this bipolar region close to the disk centre is almost identical
to the radial field component on the solar surface.
For this reason we can employ here only a linear force-free field model
for the extrapolation of the magnetogram
\citep{seehafer:1978}.

\section{The reconstruction}

\begin{figure*}
  \hspace*{\fill}
  \includegraphics[height=7cm]{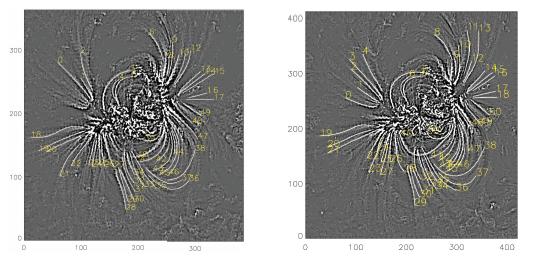}
  \hspace*{\fill}
  \hspace*{\fill}
  \caption{Contrast enhanced zoom of the EUVI images B (left) and
    A (right) of the active region NOAA 0960. Heliographic north
    is upward. The axes are scaled according to the image pixel size.
    Individual loop structures are emphasized by white curves and
    enumerated. Equal numbers do not imply a correspondence
    across the images.}
  \label{fig:loopAB}
\end{figure*}

The first step in the stereoscopic reconstruction scheme is the
isolation and identification of individual loops in each of the EUV
images.
In Figure~\ref{fig:loopAB} we show the portion of the EUV images containing
the active region. The EUV structures were contrast enhanced by an
unsharp mask filter. Next, individual loop structures were detected by a
loop segmentation program. This program detects individual bright
loops in an image by treating them as elongated intensity ridges
\citep{inhester:etal:2007}.
For identification, the loop curves were enumerated. These assignments,
e.g. a number $i_A$ for a loop curve in image A, were made
independently in each image.

\begin{figure}
  \hspace*{\fill}
  \includegraphics[height=6cm]{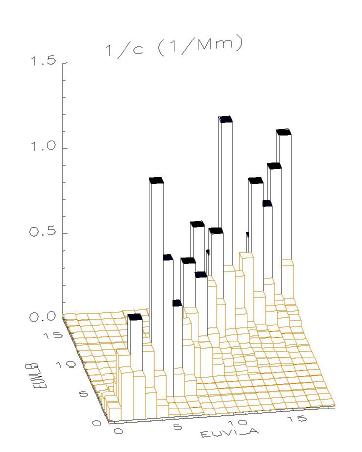}
  \hspace*{\fill}
  \caption{Proximity of loops identified in images A and B in
    figure~\ref{fig:loopAB}. The proximity is expressed by the
    inverse of a distance measure $C$ (see text).
    The loops from image A and B are arranged along axes `EUVI\_A'
    and `EUVI\_B' according to their respective identification number
    $i_A$ and $i_B$.
    For each pair $(i_A,i_B)$ the inverse of $C$ is displayed by a
    column at the location of the loop pair in this matrix
    representation. Columns exceeding 0.5 Mm$^{-1}$
    have a black top. Here, only loops from the northern
    half of the active region were considered.}
  \label{fig:correspupper}
\end{figure}

To establish correspondences of projections $i_A \rightleftarrows i_B$
of the same loop across the images is the hardest part in the
stereoscopy procedure. For isolated loops they can sometimes be
guessed by visual comparison of the image pair.
Also, some guidance is provided by matching constraints which
corresponding pairs of loop projections have to obey \citep{inhester:2006}.
Often, however, the visual comparison of loop structures
does not yield unique correspondences.
To disentangle the typically crowded active region loop ensembles we
have developed a systematic scheme which determines correspondences
with the help of magnetic field model calculations
\citep{wiegelmann:etal:2005,wiegelmann:inhester:2006,feng:etal:2007}.
The idea is to find three dimensional field lines from a
more or less accurate model of the active region magnetic field as
a first approximation to the final loops whose projections are close
to the loop projections identified in the images from spacecraft A and B.
If a field line can be found with projections sufficiently close to a
loop in both images, this is strong evidence that these loop curves
represent projections of the same three-dimensional loop.

We quantify the proximity of a projected field line $l$ to a loop
curve $i_A$ in image A, say, by the mean distance $C_A(i_A,l)$ between
the two-dimensional curves in this image. The probability of a
correspondence between a pair $(i_A,i_B)$ of loop curves in image A and
B can then be measured by
$C = \half\mathrm{min}_l\,(C_A(i_A,l)+C_B(i_B,l))$.
Here, the set of possible field lines $l$ comprised
all possible foot point locations and a wide range of $\alpha$ values
from -0.01 to +0.01 Mm$^{-1}$.
The field lines $l$ here only serve as a means to
establish the correspondence, they are not intended to represent a
consistent field model of the active region.
The linear force-free field model used is only consistent if $\alpha$ is a
global constant. Strictly speaking, the field lines $l_\mathrm{min}$
for which $C$ attains the minimum are each from a different field line
model as $\alpha$ turned out to differ for each loop pair.

In Figure~\ref{fig:correspupper} the inverse of $C$ is shown for
the loops in the northern half of the active region,
$i_A$ = 0 to 18 and $i_B$ = 0 to 17.
Some few loop combinations show a clearly enhanced $1/C$ and are thus
much more probable than the majority of combinations $(i_A,i_B)$.
We accepted for a reconstruction only loop pairs with a value of $C$
below 2 Mm.
This corresponds to an average distance between the
field line projection and the loop curves in each image of 2 pixels
or less.
When more than one combination was possible for one loop, the most
probable one was taken such that each loop receives no more than one
partner and the sum of $C$ of all selected correspondences was
minimised \citep{wiegelmann:inhester:2006}.
In all, 20 pairs from Figure~\ref{fig:loopAB}
could thus be identified.

The last step is the stereoscopic reconstruction of the
three-dimensional loop from each accepted pair $(i_A,i_B)$. This
purely geometrical step often yields multiple solutions
\citep{inhester:2006}. They were discarded by retaining only
the three-dimensional reconstruction closest to the
best fit field line $l_\mathrm{min}$.

\section{Results}

\begin{figure}
  \hspace*{\fill}
  \includegraphics[width=7cm]{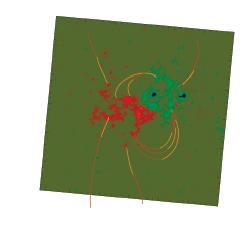}
  \hspace*{\fill}
  \caption{Vertical View of the three-dimensional reconstruction results
           from a viewpoint within a degree from the STEREO A spacecraft.
           Heliographic north is upward. The
           reconstructed loop sections are drawn in yellow, the closest
           fit field lines in red. The loop pairs $(i_A,i_B)$ drawn are:
           4-2, 12-12, 5-3, 7-5 (northward part of this AR) and
           45-45, 44-43, 42-42, 24-23, 30-29 (southward part).}
  \label{fig:3dloopline4}
\end{figure}

\begin{figure}
  \hspace*{\fill}
  \includegraphics[width=9cm]{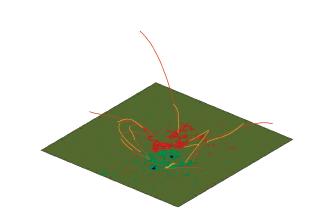}
  \hspace*{\fill}
  \caption{Same as Figure~\ref{fig:3dloopline4}, but seen from a view point
           NE of the active region. Heliographic north points to the lower
           left corner.
           The SECCHI instruments observed from approximately above.}
  \label{fig:3dloopline3}
\end{figure}

In Figures~\ref{fig:3dloopline4} and \ref{fig:3dloopline3} we present
two views of a set of reconstructed loops.
Figure~\ref{fig:3dloopline4} shows the reconstructed loops (yellow)
and the associated closest fit field lines(red) obtained by extrapolation
from a position within a degree from the STEREO A spacecraft.
As expected, loops and field lines agree relatively well from this perspective
because they were chosen to be close in this projection.
Figure~\ref{fig:3dloopline3} therefore provides a completely different view
of the active region.
This view shows that most of the loops cannot easily be approximated
by planar curve segments. This geometrical simplification was often
used for loop reconstructions in the past because a more involved shape
could only rarely be resolved from previous observations.
This figure also reveals deviations between the loops
and field lines.
E.g., the loops on presumably open field lines appear to be more
strongly curved than the corresponding field lines from the
extrapolation.

\begin{figure}[t]
  \hspace*{\fill}
  \includegraphics[width=9cm]{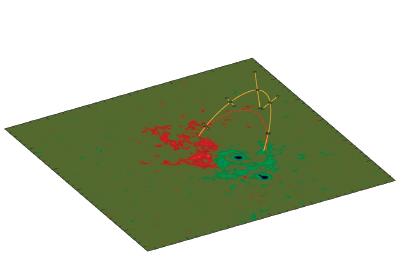}
  \hspace*{\fill}
  \caption{Example of a reconstructed loop with error estimates.
           The reconstruction is shown in yellow along with error bars.
           The associated best-fit linear force-free field line (red curve)
           is much lower in height.}
  \label{fig:3dloop5A3B_e}
\end{figure}

We attribute this disagreement to a deficiency of the linear
force-free field line extrapolation.
For the closed field lines, the best fit $|\alpha|$ values derived
above fell in the range from 1.8 to 8.3 10$^{-3}$ Mm$^{-1}$ (see table
\ref{tab:LoopParam}).
For the open field lines, these values turned out to be
smaller in magnitude, with values $|\alpha|<$2.5 10$^{-3}$ Mm$^{-1}$.
As $\alpha$ is a global constant for the linear force-free field model,
the influence of the stronger currents on the closed active-region field
lines is not accounted for on the open field lines. This may explain why
the open field lines were calculated with less curvature than the
corresponding stereoscopically reconstructed loops.

\begin{table}
\hspace*{\fill}
\begin{tabular}{ccccc}
\hline
Loop pair & $|\alpha|$ &height &length \\
$i_A,\,i_B$ &($10^{-3}Mm^{-1}$)& (Mm) & (Mm) \\ \hline
\hline
 5,\hspace{1em}3   &1.8   &71.9  & 229 \\
 7,\hspace{1em}5   &8.3   &20.6  & 105 \\
 45,45             &2.3   &58.2  & 253 \\
 44,43             &2.8   &27.3  & 188 \\
 42,42             &2.8   &57.2  & 210 \\ \hline
\end{tabular}
\hspace*{\fill}
\caption{Best fit field line parameters for a
 representative list of closed loops of active region NOAA 0960}
\label{tab:LoopParam}
\end{table}

The loop reconstruction is also prone to errors, however. These may occur
whenever a projected loop section in the
images are directed tangentially to an epipolar line
\citep{inhester:2006}.
For the viewing geometry of our observations, epipolar lines are
nearly horizontal in the images and the critical part for closed, E-W
orientated loops therefore lies more or less near their apex.
Also the open loop structures 16-19 in image B and 17-20 in image A
(see Figure~\ref{fig:loopAB}) suffer from this problem as they are
orientated almost entirely horizontally in the images. We have therefore
not attempted to reconstruct them even though a correspondence could
well be identified.

In Figure~\ref{fig:3dloop5A3B_e} we display the reconstruction of loop
(5,3) (yellow curve) which shows by far the largest deviation to its
best fit linear force-free field line (red curve). For most other
loops, this discrepancy is much less although the agreement is rarely
perfect. For some points along the loop (5,3), we also show error bars
which represent the geometrical reconstruction error when the
uncertainty for the loop projection in the images is assumed to be 1.5
pixels. In this case, the height of the loop top turns out to be
$\sim$ 1.5 times above that of the corresponding field line. This
field line (the first entry in table~\ref{tab:LoopParam}) again shows
a relatively small value $|\alpha|$. Since this $\alpha$ value
gave the best fit of linear force-free field lines to the loop
projection in the images, we conclude that the linear force-free
assumption is often not adequate \citep[cf.][]{wiegelmann:etal:2005}.

\section{Discussion and outlook}

We demonstrated that EUV data from the new STEREO spacecraft allows
for the first time to make a reliable stereoscopic reconstruction of
the spatial distribution of hot, magnetically confined coronal plasma
and, by inference, provide a full three dimensional view of the
arrangement of coronal field lines.
We found that linear force-free field models are helpful to establish
correspondences between the loops observed in the STEREO image pairs.
The field lines from these linear force-free models need not be physical
but only serve as a first order approximation to the
final loops.
Realistic magnetic field models of the corona will
have to be judged by their capability to yield field lines
in agreement with the stereoscopically reconstructed loops.
Our scheme to determine correspondences will become even
more valuable when the stereo base angle grows and loop structures become
more difficult to be identified in the image pairs.

The reconstructions will also allow more precise analyses of emissions from
loops. The observed brightness of EUV loops is, e.g., strongly modified
by the inverse cosine of the angle between the line of sight and the loop's
local tangent. This may, besides other effects,
contribute to the enhanced EUV brightness of the lower loop segments
commonly observed on the solar disk: these loop segments close to the
loop's foot points are more aligned with the radial direction and they make a
small angle with the view direction. This may cause them to appear brighter
than the loop top which is viewed at more or less right angles.

Other applications have been proposed
\citep{aschwanden:2005,aschwanden:etal:2006}. E.g., the amount of
twist of a reconstructed loop indicates how close the flux tube is to
a kink instability. \citet{Torok:etal:2004} found a threshold of about
$3.5\pi$ in numerical simulations for the twist
$\Phi=LB_\phi/rB_{\parallel}$. Here $L$ is the length of the flux tube,
$B_{\parallel}$ the toroidal field along its axis and $B_\phi \simeq
\alpha B_{\parallel} r/2$ the poloidal field at a radius $r$ from the
flux tube centre. In some cases it may be possible to resolve the
number of turns $n$ which a field line makes about the flux tube
centre from stereoscopic reconstruction
and thus to determine the twist
from $\Phi=2\pi n$. Likewise, the twist is also related to $\alpha$ and
$L$ by $\Phi=\alpha L/2$. For the active region observed here,
table~\ref{tab:LoopParam} gives values of $\Phi < 0.5$ well below the
kink instability threshold.

Another perspective for stereoscopic loop reconstruction is
the analysis of loop oscillations from a series of image
pairs. The reconstructed loops will allow us to determine the
transverse polarisation of these oscillations
\citep{aschwanden:etal:2002,Wang:Solanki:2004}. Since the coronal
magnetic field has a complicated geometry without symmetries, the
frequency of these oscillations will
significantly depend on this polarisation. Note that these
phenomena are invisible in the magnetic surface data and therefore
cannot be retrieved from field extrapolations, which in addition
require a stationary magnetic field.

\acknowledgements

LF was supported by the IMPRS graduate school run jointly
by the Max Planck Society and the Universities G\"ottingen and Braunschweig.
The work was also supported by DLR grant 50OC0501.

The authors thank the MDI/SOHO and the SECCHI/STEREO consortia for the
supply of their data.
STEREO is a project of NASA, SOHO a joint ESA/NASA project.
The SECCHI data used here were produced by an
international consortium of the Naval Research Laboratory (USA),
Lockheed Martin Solar and Astrophysics Lab (USA), NASA Goddard Space
Flight Center (USA), Rutherford Appleton Laboratory (UK), University of
Birmingham (UK), Max-Planck-Institut for Solar System Research (Germany),
Centre Spatiale de Li\`ege (Belgium), Institut d'Optique Th\'eorique et
Applique\'e (France), Institut d'Astrophysique Spatiale (France).


\end{document}